\documentclass[useAMS,usenatbib]{mn2e}

\usepackage{graphicx}
\usepackage{psfig}

\newcommand\aj{AJ}
\newcommand\apj{ApJ}
\newcommand\apjl{ApJ}
\newcommand\apjs{ApJS}
\newcommand\aap{A$\&$A}
\newcommand\mnras{MNRAS}
\newcommand\prd{Phys.~Rev.~D}
\newcommand{\bm}[1]{{\mbox{\boldmath $#1$}}}

\title[Reconstructing distributions from photo-$z$s]{Reconstructing galaxy fundamental distributions and scaling
       relations from photometric redshift surveys. Applications to the SDSS early-type sample}

\author[Graziano Rossi et al.]
{Graziano Rossi$^{1}$\thanks{Email: graziano@kias.re.kr}, Ravi K. Sheth$^{2}$ and Changbom Park$^{1}$\\
\\
$^{1}$ Korea Institute for Advanced Study, Hoegiro 87, Dongdaemun-Gu, Seoul $130-722$, Korea\\
$^{2}$ Department of Physics and Astronomy, University of Pennsylvania, 209 South $33^{rd}$ Street,
       Philadelphia, PA $19104-6396$, USA}
\date{\today}

\pagerange{\pageref{firstpage}--\pageref{lastpage}}
\pubyear{0000}

\begin{document}
\maketitle
\label{firstpage}



\begin{abstract}

Noisy distance estimates associated with photometric rather than
spectroscopic redshifts lead to a biased estimate of the
luminosity distribution, and produce a correlated mis-estimate of the sizes.
We consider a sample of early-type galaxies from the SDSS
DR6 for which both spectroscopic and photometric information is
available, and apply the generalization of the $V_{\rm max}$ method 
to correct for these biases. We show that our technique recovers
the true redshift, magnitude and size
distributions, as well as the true size-luminosity relation.
We find that using only $10\%$ of the spectroscopic information randomly spaced
in our catalog is sufficient for the reconstructions to be accurate
within $\sim 3\%$, when the photometric redshift error is $\delta z
\simeq 0.038$. 
We then address the problem of extending our method to deep redshift catalogs, 
where only photometric information is available.
In addition to the specific applications outlined here, our technique
impacts a broader range of studies, when at least
one distance-dependent quantity is involved. It is particularly
relevant for the next generation of surveys, some of which will only
have photometric information. 

\end{abstract}



\begin{keywords}
distance scale -- galaxies: distances and redshifts -- methods: statistical -- galaxies: formation --- catalogues -- survey
-- galaxies: fundamental parameters -- cosmology: observations.
\end{keywords}



\section{Introduction}


The redshift and luminosity distributions of galaxies, 
and galaxy scaling relations, such as the color-magnitude relation, the
size-surface brightness relation, the
luminosity-size relation or the Fundamental Plane, play a
crucial role in constraining galaxy formation models.
However, a bias will be intrinsically present in all these correlations
if the transformation from observable to physical quantity involves
one or more distance-dependent observables, due to noise in the
distance estimate.
Distances are only known approximately if photometric redshifts are
available, but spectroscopic redshifts are not. 
This is already the case of many current surveys (e.g. SDSS, COMBO-17, 
MUSYC, COSMOS, CFHTLS), where the number of objects with photometric redshifts is more than an order of magnitude bigger than
that of spectroscopic redshifts, and will be increasingly true of the next generations of deep
multicolor wide-area photometric surveys (e.g. DES, LSST, SNAP, JDEM), which will increase 
the number of galaxies with multi-band photometry to a few billions.

Photometric information is
essential and statistically more significant for studying cosmological evolution at a fraction of the
cost of a full spectroscopic survey.
Therefore, many efforts are currently devoted to improve photometric redshift
estimations (see for example Feldmann et al. 2006; Carliles et
al. 2008; Hildebrandt et al. 2008; Oyaizu et
al. 2008a,b; Stabenau et al. 2008; Budavari 2009; Ilbert et
al. 2009; Jouvel et al. 2009; Salvato et al. 2009), especially because accurate photometric redshifts are among
the key requirements for precision weak lensing measurements
(Banerji et al. 2008; Ma \& Bernstein 2008; Mandelbaum et al. 2008). 
Well-understood photometric redshifts and errors are also vital in
resolving redshift ambiguities where spectroscopy shows only a single
spectral line (Lilly et al. 2007), and are especially crucial to dark
energy science (Bernstein \& Huterer 2009; Sun et al. 2009).

Hence, methods for recovering unbiased estimates of the redshift distribution
(Padmanabhan et al. 2005; Sheth 2007; Lima et al. 2008),
the luminosity function (Sheth 2007), and
galaxy scaling relations (Rossi \& Sheth 2008)
from magnitude limited photometric redshift
datasets are indeed necessary. 
In particular, in Rossi \& Sheth (2008) we described
two techniques which can handle this complication (i.e. a non-parametric
deconvolution method and a maximum likelihood approach), and the
extension of the $V_{\rm max}$ algorithm (Lucy 1974) was
tested on a mock catalog.
Here we apply the same method to the SDSS DR6, and
investigate the bias present in the fundamental distributions
and in the luminosity-size relation for
early-type galaxies, which arises when computing these relations from photometric
data. The technique is insensitive to the actual
quality of photo-$z$ estimates, but it relies on the knowledge of the
conditional probability $p(z_{\rm photo}|z_{\rm spectro})$.  
In essence, if photo-$z$ errors are
at least known, our technique is applicable.

We have two main goals in this study. The first is to 
use a selected sample of early types from the SDSS DR6, for which both photo-$z$s and
spectro-$z$s are known, and apply our deconvolution 
technique to derive the unbiased redshift, magnitude and size
distributions, and the
magnitude-size relation. We refer to this procedure as the
``calibration'' part. 
The second and more challenging goal is to use our calibration in
order to infer information when spectroscopic data is poor or not available
(i.e. deep redshift catalogs). 


The outline of the paper is as follows.
Section 2 describes the SDSS galaxy catalog used in this analysis, and
highlights the criteria adopted for the early-type selection.
Section 3 presents the reconstruction of the redshift, magnitude,
size, and size-magnitude distributions
from photometric data, for the early-type sample. A brief summary of 
the deconvolution method is provided -- while we point out in an appendix
the relation between our deconvolution procedure and a
convolution-based approach --, the dependence of
$p(z_{\rm photo}|z_{\rm spectro})$ on magnitude is also discussed, as well as
other technical details. 
Section 4 deals with extending our technique when
spectroscopic information is poor, or when only
photometry is available. Some tests are performed on the early-type
``calibration'' catalog, and in particular it is found that, by using only $10\%$ of the spectroscopic
information randomly spaced in redshift space, one can reconstruct accurately
the galaxy fundamental distributions.
Finally, Section 5 summarizes our findings, 
and indicate ongoing and future studies and applications. 


Whenever necessary, we assume a spatially flat cosmological model
with $(\Omega_{\rm M}, \Omega_{\rm \Lambda},h) = (0.3, 0.7, 0.7)$, where
$\Omega_{\rm M}$ and $\Omega_{\rm \Lambda}$ are the present day densities of
matter and cosmological constant scaled to the critical density, and
write the Hubble constant as $H_0 = 100~h~$ km s$^{-1}$ Mpc$^{-1}$. 



\section{The SDSS Early-Type Sample}

The catalog we use is based on the Sloan Digital Sky Survey (SDSS)
Data Release 6 (DR6, $http://www.sdss.org/$), available online through the
Catalog Archive Server Jobs System (CasJobs).
We adopt selection criteria suitable to early-type galaxies, as
described in Bernardi et al. (2003). 
Specifically, from the DR6 galaxy photometric sample (PhotoObjAll in
the Galaxy view, which contains primary objects that are classified as galaxies), we
select objects according to these general criteria:
\begin{itemize}
\item Petrosian magnitudes in the range $14.50 \le m \le 17.45$ for the $r$ band.
\item Concentration index $R_{\rm petro,90}/R_{\rm petro,50} > 2.5$ in the $i$ band.
\item Likelihood of the de Vaucouleur's model $> 0.8$.
\item Objects with both photometric and spectroscopic redshifts available.
\end{itemize}
No redshift or velocity dispersion cuts were made, although we tested
the effect of a velocity dispersion cut 
($\sigma > 0$, so good S/N) and found no substantial difference. 
Our catalog contains 163,718 objects, and consists of model
magnitudes, petrosian radii, de Vaucouleurs and
exponential fit scale radii along with their corresponding axis ratios in the
$r$ band, photometric and spectroscopic redshifts and their quoted errors.

Model magnitudes are obtained by measuring galaxy fluxes
through equivalent apertures in all bands, and by fitting the
exponential or de Vaucouleurs model of higher likelihood in the $r$ filter
and applying it in the other bands, after convolution with a PSF in each band (for more
details, see Blanton et al. 2003). 
The previous fitting procedures yield also the effective radii of the
models and the axis ratio of the best fit model. 
In particular, the Petrosian ratio $R_{\rm P}$ at a radius $r$ from the center of an
object is defined to be the ratio of the local surface brightness in an annulus
at $r$ to the mean surface brightness within $r$, as described by Blanton
et al. (2001) and by Yasuda et al. (2001).
The Petrosian radius $r_{\rm P}$ is the radius at which $R_{\rm P}(r_{\rm P})$
equals some specified value $R_{\rm P,lim}$, set to $0.2$ in our case.  

We select photometric redshifts from the SDSS Photoz table.
This set of photometric redshifts has been obtained with the template
fitting method, which simply compares the
expected colors of a galaxy (derived from template spectral energy
distributions) with those observed for an individual galaxy. 
The empirical templates of Coleman, Wu \& Weedman (1980), extended with spectral synthesis models,
are used. These templates were adjusted to fit the
calibrations, as explained in Budavari et al. (2000).
More detailed information about the photo-$z$ catalog used here is also provided in Csabai et
al. (2003), and references therein.  
The main advantage of this technique in computing photo-$z$s
is a broader redshift range coverage  
for all types of galaxies, and the additional information like spectral type, 
K-correction and absolute magnitudes. However, 
its accuracy is severely limited by the lack of perfect spectral
energy distribution (SED) models. In fact,
the quality of photometric redshift
estimation of faint objects (or with large photometric errors) is weak.
More generally, the standard scenario for template fitting is to take a small number of
spectral templates and choose the best fit by optimizing the
likelihood of the fit as a function of redshift, type and
luminosity. Variations of this approach have been developed in the
last few decades. For example, in the recent SDSS DR7 photometric redshifts
are obtained with a hybrid method, namely a combination of the template
fitting procedure and of a technique which compares the observed colors of galaxies to a reference set that has both 
colors and spectroscopic redshifts observed. 

\begin{figure}
\centering
\includegraphics[angle=0,width=0.5\textwidth]{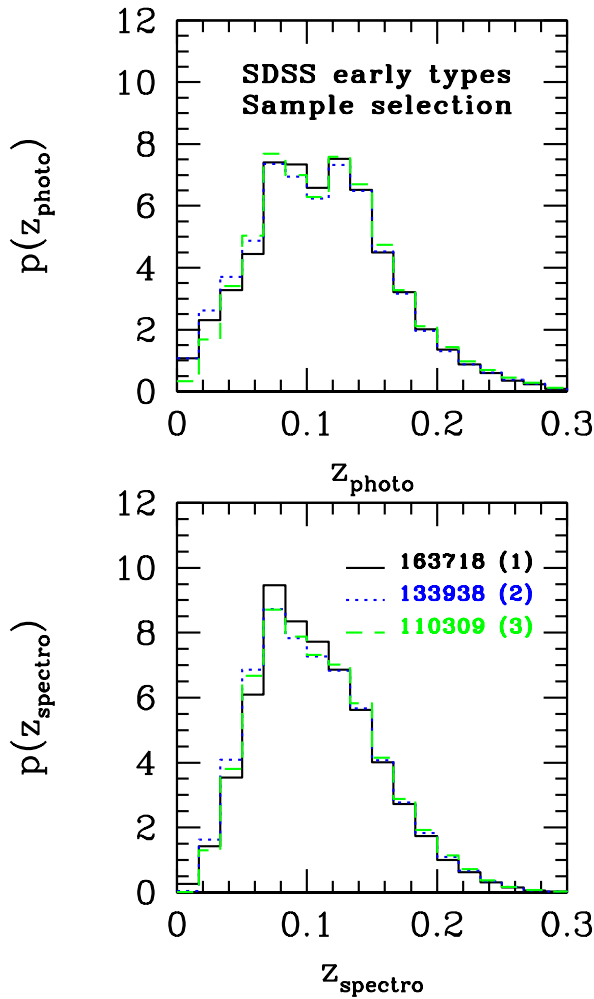}
\caption[Photometric and spectroscopic SDSS redshift distributions for
  the early-type calibration catalog: 
effect of the selection criteria]
{Photometric and spectroscopic 
     SDSS redshift distributions for the early-type sample: effect
  of the selection criteria. Solid lines represent the redshift
  distributions of the sample considered in this study (Sample 1). Dotted lines
  are the result of using only spectra of good quality (Sample 2). Dashed
  lines denote a more sophisticated selection process, as explained in
  the main text (Sample 3).}
\label{redshift_cal_comparisons_early_types}
\end{figure}

\begin{table}
 \caption[]{Median spectroscopic and photometric redshifts, 
corresponding median absolute deviations (MAD),  
standard normalized median absolute deviations
(NMAD) and dispersions for the three early-type samples illustrated in Figure \ref{redshift_cal_comparisons_early_types}.
}
\begin{center}
\begin{tabular}{|l|c|c|c|}
\hline\hline
  &SAMPLE 1& SAMPLE 2& SAMPLE 3 \\
\hline\hline
median $z_{\rm spectro}$& 0.1021& 0.1021& 0.1040\\
median $z_{\rm photo}$& 0.1107& 0.1093& 0.1131\\
MAD $z_{\rm spectro}$& 0.0307&  0.0328&  0.0327\\
MAD $z_{\rm photo}$& 0.0348& 0.0361& 0.0358\\
NMAD&  0.0172&  0.0173&  0.0154\\
$\sigma_{\rm \Delta z /(1+z_{\rm spectro})}$& 0.0340& 0.0263& 0.0219\\
\hline
\end{tabular}
\end{center}
\label{fig_1_info}
\end{table}

We finally cross-correlate the photometric information with the SDSS DR6
spectroscopic sample (SpecObjAll), and select only those photometric
objects for which spectroscopic information is also available.
The spectroscopic pipeline (spectro1d) assigns a final redshift to each object
spectrum by choosing the emission or cross-correlation redshift with
the highest likelihood and stores this as $z$ in the specObj table. In addition
to spectral classification based on measured lines, galaxies are
classified by a Principal Component Analysis (PCA), using cross
correlation with eigentemplates constructed from the SDSS spectroscopic
data. In the selection of our sample we use only photometric criteria,
but more robust constraints can be applied in order to reduce
galaxy-type errors, and their effect is illustrated in Figure \ref{redshift_cal_comparisons_early_types}.
In both panels, solid lines represent the redshift distributions of the
calibration sample used in this study (Sample 1).
However, if in addition we require spectra of good quality or without masked regions (SDSS
warning flag for spectra of low quality set to zero),
then the number of galaxies drops to $133,938$ (dotted lines in
Figure \ref{redshift_cal_comparisons_early_types}, Sample 2).
Finally, if we consider only spectra with PCA classification numbers
$a<-0.1$, typical of early-type galaxy spectra (Connolly \& Szalay
1999), we find $110,309$ objects (dashed lines in
Figure \ref{redshift_cal_comparisons_early_types}, Sample 3).
For all these samples, we provide in Table \ref{fig_1_info} 
the median spectroscopic and photometric redshifts, their 
corresponding median absolute deviations (MAD), the  
standard normalized median absolute deviation
(NMAD) defined as in Hoaglin et al. (1983) by
$1.48 \times median[|\Delta z|/(1+z_{\rm
spectro})]$, and the dispersion $\sigma_{\rm \Delta z /(1+z_{\rm spectro})}$, where
$\Delta z = z_{\rm spectro} - z_{\rm photo}$.

In our study we consider the sample obtained with the photometric-only selection
process (Sample 1). This is because our second goal is to rely on this ``calibration''
subset to infer information when only photometry is available.  

More sophisticated criteria for obtaining a well-controlled sample of early-type
galaxies are presented in Park and Choi (2005) who used color,
color-gradient, and concentration index to classify galaxies into
early and late types with reliability and completeness exceeding $90\%$.
See also Hyde \& Bernardi (2009), where problems like
contamination by later-type galaxies and
systematic effects due to the use of Petrosian quantities are
addressed in detail. However, since we are not attempting to make a precision measurement of scaling
relations, but rather our main goal is to show how to correct for
photo-$z$ biases, more robust selection criteria do not affect the nature
of the problems investigated in this study.



\section{Deconvolutions from photometric data} \label{reconstr}

In this section we apply our reconstruction technique based on the
generalization of the $V_{\rm max}$ method (Sheth 2007; Rossi \& Sheth 2008) and
briefly summarized here to the redshift, magnitude and size
distributions, and to the size-luminosity relation of the early-type sample.
A new deconvolution code named \textit{DeFaST} (acronym for \textit{Deconvolution Fast}, with the convention of using
capital letters for consonants), which performs a fast integral
deconvolution, has been developed for this study. Lucy's (1974) iterative scheme is implemented, in one
or two dimensions. Appropriate variations have been carried out in order to handle correctly
different choices of the conditional probability functions. In
particular, for the SDSS early-type sample the conditional distributions are measured directly from the data and then
used in the deconvolution algorithm. Splines fits to the
pdf's are performed in those cases. 
For technical details we refer the reader to a trial
version of the one-dimensional software, freely available for download
online at the web address $http://www.physics.upenn.edu/\sim grossi/research/software.htm$.


\subsection{Extended $V_{\rm max}$ method as a non-parametric deconvolution-like technique} \label{dm}

The $V_{\rm max}$ method, originally devised by Schmidt (1968), is a
way of testing the uniformity of spatial distributions.  
In particular, if $V$ is the comoving volume between an object in
a flux-limited catalog at redshift $z$ and the
observer located at $z=0$, and $V_{\rm max}$ is the corresponding total survey volume over which the
same object could have been seen at $z_{\rm max}$, then clearly the
ratio $V/V_{\rm max}$ is a measure of the position of the source. Therefore, if a
distribution is uniform then the average value over all the objects in the catalog, $\langle V/V_{\rm max}
\rangle$, must be 0.5. 
The $V_{\rm max}$ method allows one to correct for selection effects
present in magnitude-limited
datasets, where fainter objects are seen
only at closer distances. In fact, in order to properly estimate for
example the luminosity function one must sum over all the objects in
the catalog and weight each source separately by the inverse of
$V_{\rm max}$ (or the inverse of $V_{\rm max}-V_{\rm min}$ if the
catalog is limited at both ends). This is the basis of the procedure
developed by Schmidt (1968). 

Generalizations of this method to include distance errors
have been carried out in Sheth (2007) for the luminosity function (1D case), and
in Rossi \& Sheth (2008) for galaxy scaling relations (2D case, or
full $n$-dimensional manifold).
To summarize, in a flux-limited survey the quantities affected by the
photometric redshift errors are the intrinsic luminosity distribution
rather than the luminosity function itself (which differs from the
previous one by the
inclusion of a $1/V_{\rm max}$ weighting), 
and the intrinsic joint distribution of luminosities and sizes
-- or in
general the joint distribution of two (or more) observables affected by the same distance errors. 
However, in a real experiment
one measures 
their noisy counterparts.
Therefore it is necessary to reconstruct the intrinsic distributions first, before
applying the $V_{\rm max}$ prescription. 
This is achieved by recognizing 
the deconvolution nature of this class of problems, hence
an iterative algorithm is suitable for the reconstructions.

In fact, adopting Lucy's (1974) formalism, the general $n$-dimensional problem is
that of estimating the frequency distribution $\Psi({\bm \xi})$
of the intrinsic $n$-dimensional vector ${\bm \xi}$ when the available
observed measures, denoted by the vector ${\bm x}$, 
are a finite sample drawn from an infinite population characterized by
\begin{equation}
\Phi({\bm x}) = \int \Psi({\bm \xi})~p({\bm x}|{\bm \xi})~{\rm d}{\bm \xi},  
\label{Phi}
\end{equation}
where $\Phi({\bm x})$ is the data function accessible to measurements
and $p({\bm x}|{\bm \xi})$ is the conditional probability of estimating
${\bm x}$ when the true value is ${\bm \xi}$.
The iterative procedure to invert the previous expression is
\begin{equation}
 \Psi^{\rm r+1} ({\bm \xi}) = \Psi^{\rm r}({\bm \xi})
  \int {\rm d}{\bm x}~\frac{\tilde{\Phi} ({\bm x})}{\Phi^r ({\bm
  x})}~p({\bm x}|{\bm \xi}),
 \label{Psi_nd}
\end{equation}
where 
\begin{equation}
 \Phi^{\rm r}({\bm x}) = \int {\rm d}{\bm \xi}~\Psi^{\rm r}({\bm
 \xi})~p({\bm x}|{\bm \xi}).
 \label{Phi_r}
\end{equation}
The index $r$ indicates the $r$th iteration in the sequence of 
estimates, and $\tilde \Phi$ is an approximation to $\Phi$ 
obtained from the observed sample.
The starting value $\Psi^{0}(\bm \xi)$, which initializes the iteration,
should be a smooth, non-negative function having the same integrated  
density as the observed distribution. 
In our deconvolution procedure, 
we always use the observed histograms (i.e. photo-$z$ derived
distributions) as convenient starting guesses.
 
Clearly, the outlined formalism is readily applicable to the size-luminosity
correlation if we interpret ${\bm x}$ as the 2D vector of the
estimated absolute magnitudes and sizes, and 
${\bm \xi}$ as the vector of the corresponding true (or intrinsic) quantities.
Similarly for the redshift, magnitude and
size distributions, where now vectors simply reduce to scalar
quantities (1D case).  


\subsection{Redshift distribution} \label{red}

\begin{figure*}
\centering
\includegraphics[angle=0,width=0.85\textwidth]{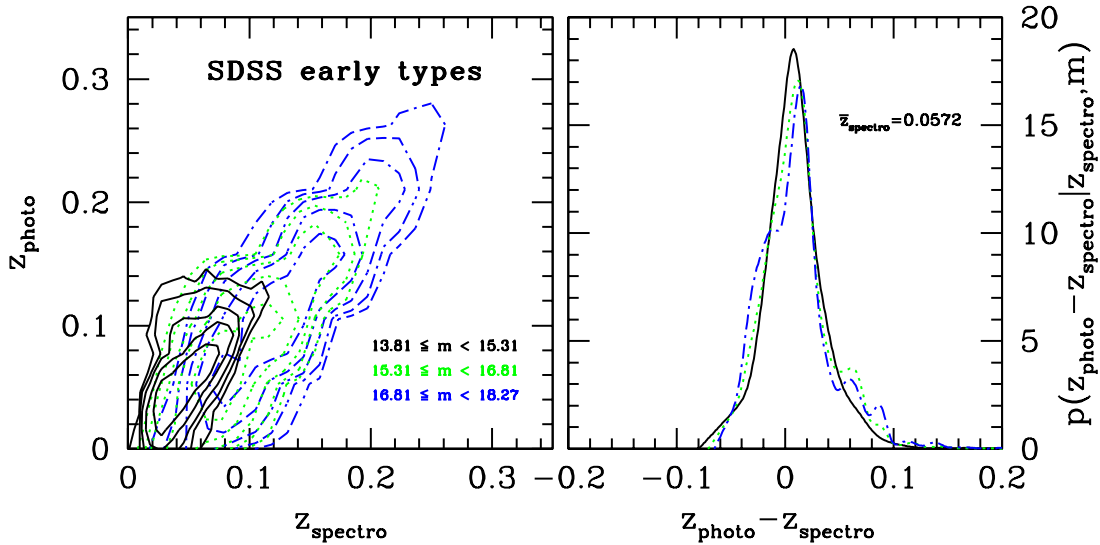}
\caption{Apparent magnitude dependence of the conditional probability $p(\zeta|z)$. [Left panel] Contours show levels
  which are $1/2^n$ times the height of the maximum value of the density
  of sources, with $n$
  running from $1$ to $5$, for the three magnitude bins specified in the figure.  
  [Right panel] Example of $p(\zeta-z|z,m)$ for $z=0.0572$ and for
  the three previous bins in magnitude.}   
\label{resdhift_magnitude_p_conditional}
\end{figure*}

Following Rossi \& Sheth (2008), we indicate with $\zeta$
and $z$ the photometric and spectroscopic redshifts, respectively.
As argued before, the problem of estimating the intrinsic redshift distribution $N(z)$
-- number of objects which lie at redshift $z$ --
is best thought of as a deconvolution
problem, and if $p(\zeta|z)$ is the probability of estimating the
redshift as $\zeta$ when the true value is $z$, then the distribution
of estimated redshifts is
${\cal N}(\zeta) =  \int N(z)~p(\zeta|z)~{\rm d}z$. 
Note that this is just a particular 1D case of equation (\ref{Phi}), where
${\bm \xi} \rightarrow z$, ${\bm x} \rightarrow \zeta$, $\Phi
\rightarrow {\cal N}$ and $\Psi \rightarrow N$.
If $p(\zeta|z)$ is known and ${\cal N}(\zeta)$ is measured, then the previous relation
is a Fredholm equation of the first kind, easily solvable with a
one-dimensional inversion algorithm. Before showing the reconstructed intrinsic redshift distribution,
we focus our attention on the conditional probability $p(\zeta|z)$. In
reality, this distribution does depend weakly on apparent magnitude. 
To illustrate the effect, we split our early-type sample in
three intervals of apparent magnitudes, approximately spaced in bins of $1.5$ magnitude
width, i.e. $13.81 \le m < 15.31$ (solid lines in Figure \ref{resdhift_magnitude_p_conditional}), 
$15.31 \le m < 16.81$ (dotted lines in Figure \ref{resdhift_magnitude_p_conditional}), 
$16.81 \le m < 18.27$ (dashed lines in Figure \ref{resdhift_magnitude_p_conditional}).
In the left panel of Figure \ref{resdhift_magnitude_p_conditional}, contours show levels
which are $1/2^n$ times the height of the maximum value of the density
of sources, with $n$
running from $1$ to $5$, for the three magnitude bins.  
In the same figure, the right panel shows an example of
$p(\zeta|z,m)$ for each of the three bins in magnitude and a
spectroscopic redshift bin centered on $z=0.0572$. 

\begin{figure*}
\centering
\includegraphics[angle=0,width=0.85\textwidth]{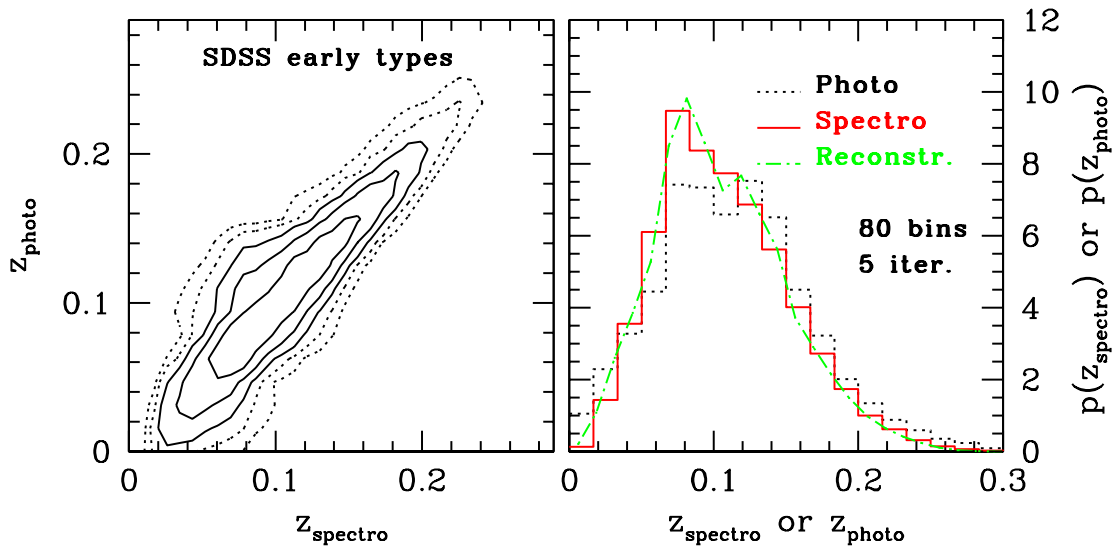}
\caption{[Left panel] Distribution of spectroscopic and photometric
          redshifts in our SDSS early-type
          catalog. Contours as in Figure \ref{resdhift_magnitude_p_conditional}.  
          [Right panel] Observed, intrinsic and reconstructed redshift
          distributions. The dotted histogram was used as a starting guess
          for the one-dimensional deconvolution algorithm. Convergence
          is achieved after a few iterations.}
\label{redshift_reconstruction_early_types}
\end{figure*}

Neglecting this small dependence of magnitude does not affect the
reconstruction of the intrinsic redshift distribution significantly.
Results are shown in Figure
\ref{redshift_reconstruction_early_types}, where in the left panel we
compare $\zeta$ and $z$, whereas in the right panel we show the
photometric or observed redshift distribution (dotted line), the
spectroscopic or intrinsic distribution (solid line) and its
reconstruction after a few iterations (dashed line), obtained by applying the one-dimensional
deconvolution algorithm based on
the Lucy's (1974) inversion technique. The error distributions used in the
reconstruction are inferred directly from the SDSS early-type data.
The median redshift of the spectroscopic sample is 
0.1021 (see Table \ref{fig_1_info}), while 
the median redshift of
the deconvolved photo-$z$ distribution is 0.1002.
This value is calculated 
as follows. We find the bin which divides the reconstructed spectroscopic
distribution in two roughly equal area parts. Within
that bin, we then interpolate with splines and provide the 
exact value of z for which the area of the reconstructed
distribution is split into two
equal parts.

Accurately characterizing $p(\zeta|z)$ is necessary for a reliable
deconvolution. After testing different methods, we have achieved good
results using cubic splines and found that simple Gaussian fits
provide unsatisfactory mapping of the conditional distributions (see
also Section \ref{gauss_prox}).

\subsection{Magnitude distribution}  \label{mag_dist}

\begin{figure*}
\centering
\includegraphics[angle=0,width=0.85\textwidth]{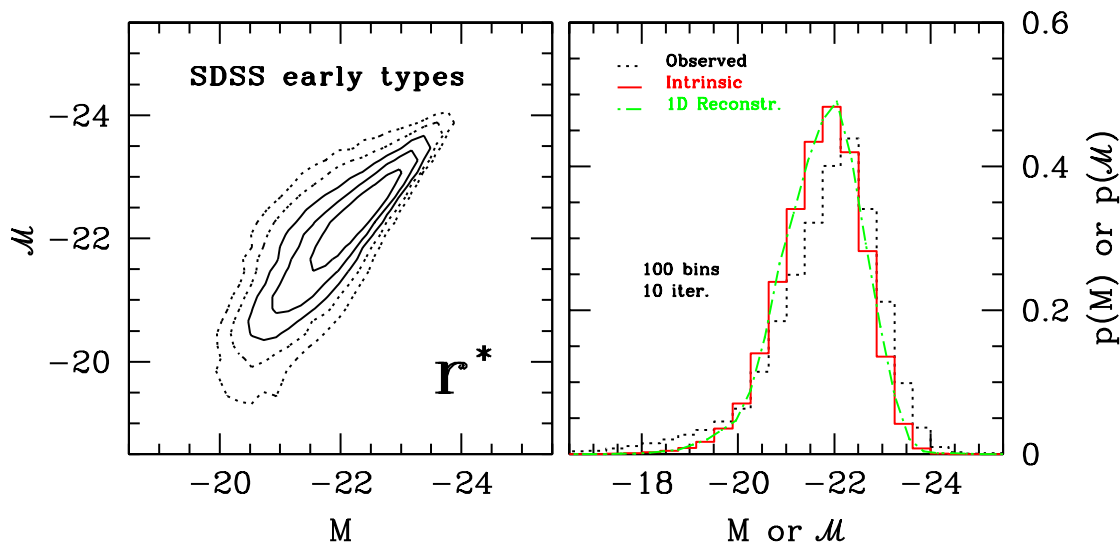}
\caption{[Left] Distributions of intrinsic and estimated absolute
  magnitudes in the SDSS early-type catalog, which result from the differences between spectroscopic
  and photometric redshifts shown in the previous figure. [Right]
  One-dimensional reconstruction of the intrinsic absolute magnitude distribution
  from the distribution of estimated redshifts. Dotted histogram shows
  the observed absolute magnitude distribution, used as a starting
  guess. Jagged line is the reconstructed intrinsic distribution, after
  $10$ iterations.}
\label{Mag_1D}
\end{figure*}

The previously outlined dependence of $p(\zeta|z)$ on apparent
magnitude suggests that one should expect, similarly, a
redshift dependence in the corresponding magnitude conditional distributions. Therefore,
it may appear difficult to  characterize and measure the appropriate conditional
probabilities, and apply the
one-dimensional deconvolution algorithm to reconstruct the magnitude
distribution.  
However the problem is simpler, as we show with the following algebra. 
Let $M$ denote the true absolute magnitude and $\cal M$
that estimated using $\zeta$ rather than $z$. Use $D_{\rm L}(z)$ to
denote the luminosity distance, and $\phi(M)$ to indicate the number density of
galaxies with absolute magnitudes $M$. Evolution is neglected. 
Let $V_{\rm max}$ denote the largest comoving volume out of
which an object of absolute magnitude $M$ can be seen, and $V_{\rm min}$ the 
analogous if the catalog is also limited at the lower end. 
The (true) number of galaxies with absolute magnitude $M$ 
(i.e. the intrinsic luminosity distribution) is:
\begin{equation}
N(M)=\phi(M)[V_{\rm max}(M)-V_{\rm min}(M)],
\label{NM_intr} 
\end{equation}
\noindent and the total number of objects with estimated absolute
magnitudes $\cal M$ is: 
\begin{eqnarray}
{\cal N}(\cal M) &=& \int {\rm d}M~\phi(M)~\Theta[V_{\rm
                 max}(M),V_{\rm min}(M),M,{\cal M}]  \\
                 &=& \int {\rm d}M~N(M)~\frac{\Theta[V_{\rm
                 max}(M),V_{\rm min}(M),M,{\cal M}]}{[V_{\rm
                 max}(M)-V_{\rm min}(M)]}, \nonumber
\label{NM_obs} 
\end{eqnarray}
\noindent where 
\begin{equation}
\Theta = \int_{D_{\rm L}(V_{\rm
    min})}^{D_{\rm L}(V_{\rm max})} {\rm d}D_{\rm L}~\frac{{\rm
    d}V_{\rm com}}{{\rm d}D_{\rm L}}~p(M-{\cal M}|M,D_{\rm L}). 
\label{Theta} 
\end{equation}
\noindent Note that since $V_{\rm max}$ and $V_{\rm min}$ are known functions
of $M$, $\Theta$ itself is just a complicated function of $M$ and
${\cal M}$.
\noindent Dividing (\ref{Theta}) by $[V_{\rm max}(M)-V_{\rm min}(M)]$ yields:

\begin{eqnarray}
\frac{\Theta}{[V_{\rm max}-V_{\rm min}]} &=& \int {\rm d}D_{\rm L}~\frac{{\rm
    d}V_{\rm com}/{\rm d}D_{\rm L}}{[V_{\rm max}-V_{\rm
    min}]}~p(M-{\cal M}|M,D_{\rm L}) \nonumber \\
    &=& \int {\rm d}D_{\rm L}~p(D_{\rm L})~p(M-{\cal M}|M,D_{\rm L}) \nonumber \\
    &=&  p(M-{\cal M}|M) \equiv p({\cal M}|M). 
\label{pM_obs_given_M_intr} 
\end{eqnarray}

\noindent Therefore, equation (5) becomes a simple
one-dimensional deconvolution, namely:
\begin{equation}
{\cal N}({\cal M}) = \int N(M)~p({\cal M}|M)~{\rm d}M.
\label{NM_obs_simpler} 
\end{equation}
\noindent The above expression (\ref{NM_obs_simpler})
is again another particular 1D case of (\ref{Phi}), where now
${\bm \xi} \rightarrow M$, ${\bm x} \rightarrow {\cal M}$, $\Phi
\rightarrow {\cal N}$ and $\Psi \rightarrow N$.
Hence, by measuring the conditional probability $p({\cal M}|M)$
from the catalog, it is possible to apply
directly the one-dimensional deconvolution algorithm -- as in the previous section. 
Note that this conclusion is particularly relevant when attempting to reconstruct
the luminosity function from photometric data.
Results of this applications are shown in Figure \ref{Mag_1D}, where
the left panel compares ${\cal M}$ and $M$, while the right   
panel shows the one-dimensional reconstruction after $10$ iterations (jagged line) of the intrinsic distribution of
absolute magnitudes (solid histogram). The observed distribution
of $\cal M$ (dotted histogram) was used as a convenient starting guess in
the deconvolution algorithm.  
We use model magnitudes in the r band, corrected for
reddening and extinction, and assume a standard cosmology
in the conversion from apparent to absolute magnitudes -- but we neglect
evolution and K-corrections. In particular, while K-corrections are necessary to properly characterize the
absolute magnitude distribution, in this paper we do not apply 
them to our de-reddened model magnitudes because 
the only purpose of our work is to describe a deconvolution technique,
which is independent of those corrections.    


\subsection{Size distribution} \label{size_dist}

\begin{figure}
\begin{center}
\includegraphics[angle=0,width=0.50\textwidth]{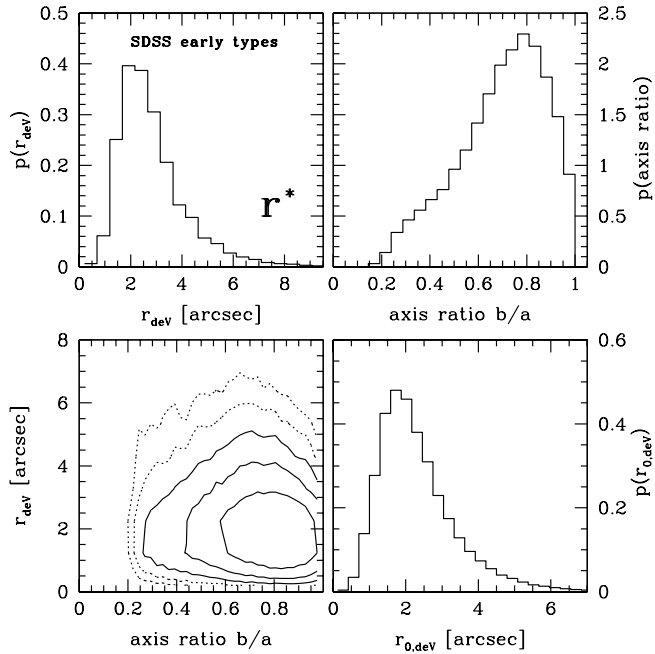}
\caption{[Upper left] Distribution of seeing-corrected
effective angular sizes of galaxies in our early-type SDSS sample. 
[Upper right] Corresponding distribution of axis ratios $b/a$.
[Lower left] Effective angular sizes $r_{\rm deV}$ as a function of
the axis ratio $b/a$. [Lower right] Distribution of effective
circular radii $r_{\rm 0, deV}$, as defined in the main text.}
\label{Size_details}
\end{center}
\end{figure}

As for the magnitude distribution, it is possible 
to reconstruct the intrinsic distribution of sizes with a one-dimensional
deconvolution algorithm (for more details, see also Section 3.3 in Rossi \& Sheth 2008). 
In what follows, we use $R$ to denote $\log_{\rm 10}$ 
of the physical size, and $\cal R$ to denote the estimated size based
on the photometric redshift $\zeta$. 
We apply one correction to convert the (seeing-corrected)
effective angular radii, $r_{\rm deV}$, output by the SDSS pipeline to physical
radii. Following Bernardi et al. (2003), we define the equivalent
circular effective radius $r_{\rm 0} = \sqrt{b/a}~r_{\rm deV}$, where $b/a$
is the corresponding axis ratio of the de Vaucouleurs radius. Then $R=\log_{\rm 10}[r_{\rm 0}D_{\rm
L}(z)/(1+z)^2]$, and similarly ${\cal R}=\log_{\rm 10}[r_{\rm 0}D_{\rm L}(\zeta)/(1+\zeta)^2]$.
We do not apply a second correction, analogous to the K-correction we
would ideally have applied to the magnitude of each galaxy, to account for the fact
that galaxies appear slightly larger in the bluer bands (i.e. Hyde
\& Bernardi 2009). 
Figure \ref{Size_details} shows the distribution of seeing-corrected
effective angular sizes of galaxies in our SDSS early-type sample (upper left panel),
the corresponding distribution of axis ratios $b/a$ (upper right
panel), the effective angular sizes $r_{\rm deV}$ as a function of
axis ratio $b/a$ (lower left panel), and the distribution of
equivalent circular effective radii $r_{\rm 0, deV}$ (lower right panel).

\begin{figure*}
\begin{center}
\includegraphics[angle=0,width=0.85\textwidth]{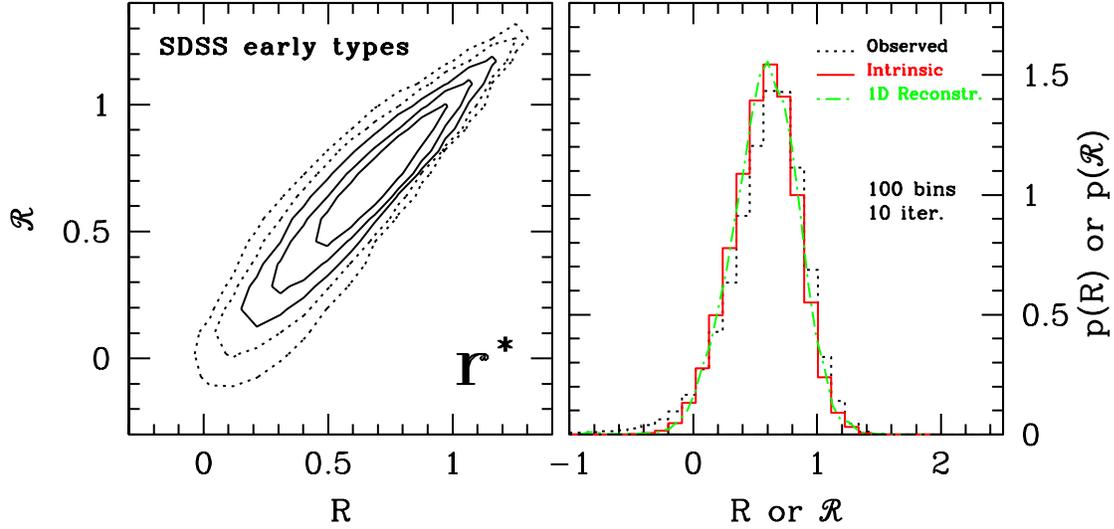}
\caption{[Left panel] Distributions of intrinsic and estimated
  physical sizes in the SDSS early-type catalog. [Right panel]
  One-dimensional reconstruction of the intrinsic size distribution from the
  distribution of estimated redshifts. Dotted histogram shows the
  observed size distribution, used as a starting
  guess. Jagged line shows the reconstructed intrinsic distribution
  after $10$ iterations. Line styles same as Figure \ref{Mag_1D}.}
\label{Size_1D}
\end{center}
\end{figure*}

In analogy with the magnitude case, we think of ${\cal N}({\cal R})$,
the number of observed objects with estimated ${\cal R}$, as being a
convolution of the true number of objects with size $R$, $N(R)$, with
the probability that an object with size $R$ is thought to have size
${\cal R}$. We measure $p({\cal R}|R)$ directly from
the catalog and run the one-dimensional deconvolution algorithm, the
result of which is presented in Figure \ref{Size_1D}.
The left panel compares ${\cal R}$ and $R$, and the right   
panel shows the one-dimensional reconstruction
(jagged line). The intrinsic distribution of
physical sizes (solid line) is recovered after a few iterations, when the observed distribution
of $\cal M$ (dotted line) is used as a convenient starting guess in
the inversion algorithm.  
Note that although the difference between the intrinsic and the observed distribution
is small, this departure will suffice to bias the
size-luminosity relation -- as we show next.

\subsection{Size-magnitude correlation} \label{size_lum_corr}

\begin{figure*}
\begin{center}
\includegraphics[angle=0,width=0.85\textwidth]{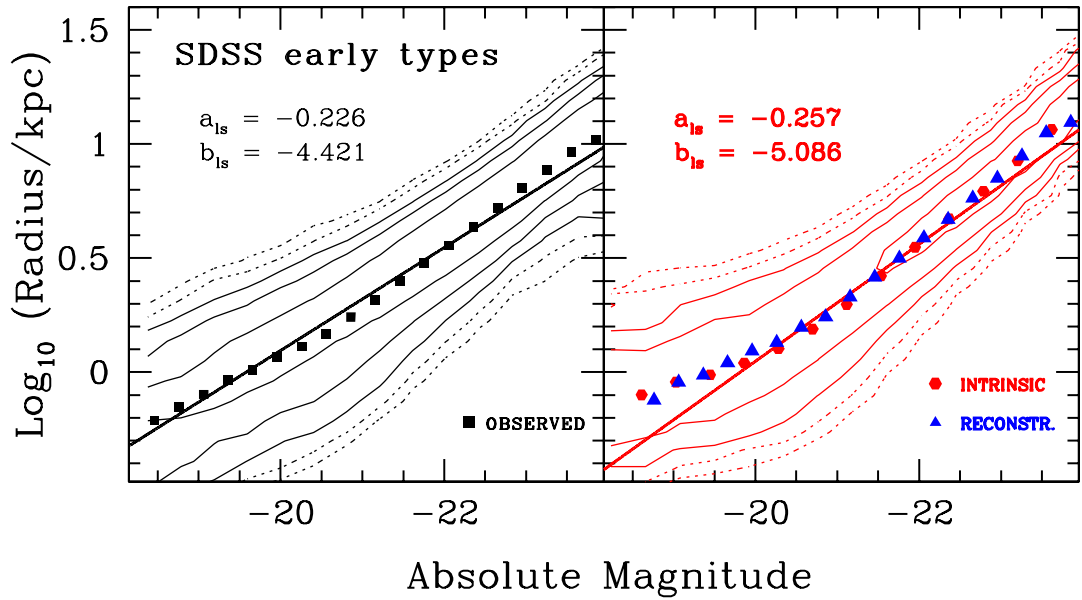}
\caption{Effect of photo-$z$s on the size-luminosity correlation in
  our SDSS early-type catalog.  In the left panel, contours and solid line
  show the ${\cal R}-{\cal M}$ relation associated with photo-$z$s,
  whereas the right panel shows the intrinsic $R-M$ relation.
  Note the bias (shallower slope in panel on left) which
  results from the fact that the photo-$z$ distance error moves
  points down and left or up and right on this plot.
  Squares in left panel show the binned starting guess for the
  2D deconvolution algorithm, triangles in right panel show
  the result of reconstruction after 7 iterations. Circles are the expected binned
  intrinsic relation, obtained from spectroscopic
  information. Convergence to the true solution is clearly seen.}
\label{RM_2D}
\end{center}
\end{figure*}

Photometric redshift errors broaden
both the magnitude and size distributions, as evident from Figures
\ref{Mag_1D} and \ref{Size_1D}, but changes to the estimated absolute
magnitudes and sizes are clearly not independent. These correlated
changes have an important effect on the size-luminosity relation, even
when the broadening of one of the two distributions is not severe. 
This is the case of our SDSS sample, where the size distribution
(Figure \ref{Size_1D}) is not severely biased, but the size-luminosity relation is still biased.
In fact, in our SDSS catalog $\langle {\cal R}|{\cal M} \rangle \propto -0.226$,
whereas $\langle R|M \rangle \propto -0.257$, as shown in Figure \ref{RM_2D}.
    
Reconstructing an unbiased estimate of size and luminosity from
photometric data is also best thought of as a non-parametric two-dimensional
deconvolution (again, see Rossi \& Sheth 2008), and application of
the extended $V_{\rm max}$ 2D algorithm to the SDSS
early-type sample is presented in Figure \ref{RM_2D}, where 
it is shown that the use of photo-$z$s 
introduces a bias in the size-luminosity relation (shallower slope in panel on left).  
Contours and solid lines indicate respectively the ${\cal R}-{\cal M}$ relation associated with
photo-$z$ (left), and the expected intrinsic $R-M$ relation (right).
Squares in left panel show the binned starting guess for the
two-dimensional deconvolution algorithm (obtained from photometric
information), triangles in right panel show
the result after $7$ iterations and circles are the expected binned
intrinsic relation, obtained from spectroscopic
information. Convergence to the true solution is clearly seen.

As pointed out in Hyde \& Bernardi (2009), most of the scaling
relations for early types show evidence for curvature. In this
respect, our technique also accounts for it, as we
reconstruct intrinsic relations within each bin, without performing any fits.  
  
\section{EXTENSIONS TO DEEP REDSHIFT CATALOGS} \label{ext}


\subsection{Challenges}

\begin{figure*}
\centering
\includegraphics[angle=0,width=0.99\textwidth]{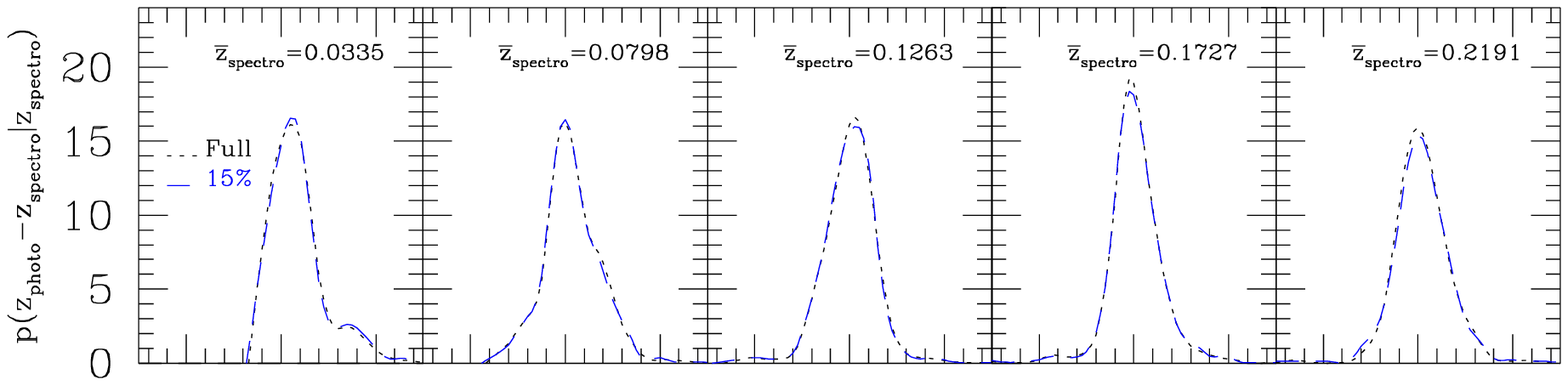}
\includegraphics[angle=0,width=0.99\textwidth]{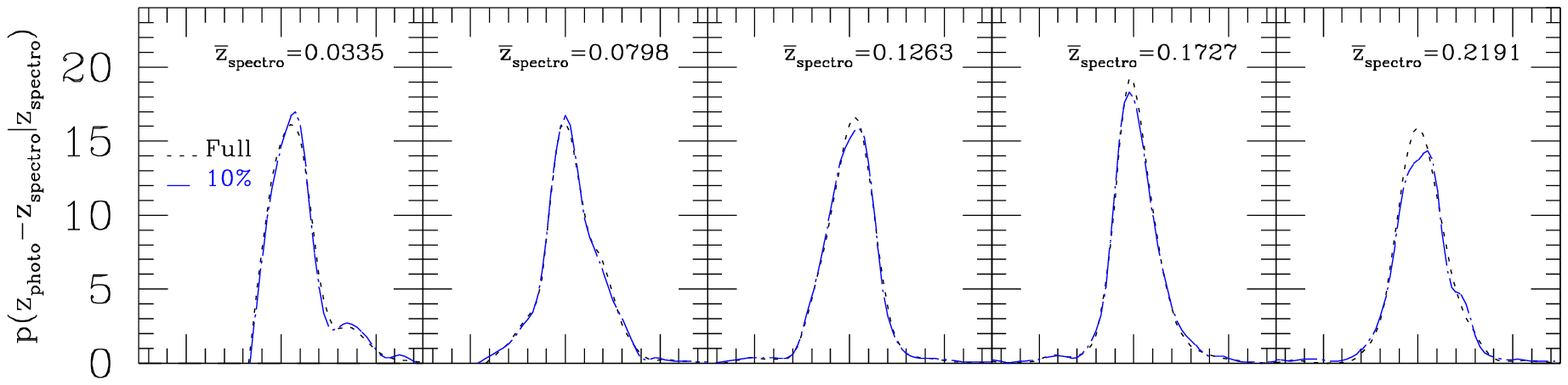}
\includegraphics[angle=0,width=0.99\textwidth]{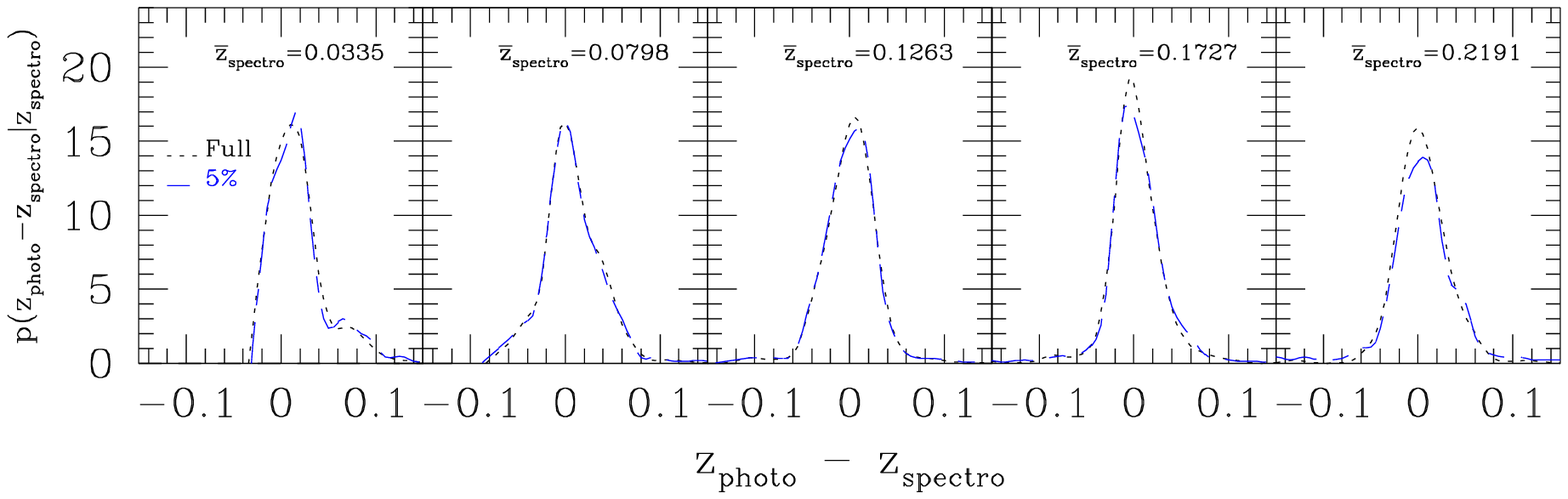}
\caption{Measurements of the conditional probabilities $p(\zeta-z|z)$'s for five different
  spectroscopic redshift bins, as indicated in the panels. Dotted
  lines are spline fits to the pdf's when the full spectroscopic
  information is used; long-dashed lines are spline fits when only
  $15\%$ [upper panel], $10\%$
  [middle panel], or $5\%$ [lower panel] of the spectral
  information is used from the original early-type catalog.} 
\label{test_number_spectra_pdf_redshift}
\end{figure*}

If we derive galaxy fundamental distributions and scaling relations using only photometry, 
a bias will be intrinsically present -- as was shown in the previous
section. With our deconvolution procedure we can
account and correct for it. However, 
our $V_{\rm max}$ reconstruction method assumes that the distribution of
photo-$z$ errors is known accurately.  
This means that spectroscopic redshifts are available for a subset of
the data, as it happens with our SDSS ``calibration'' sample.
Suppose now that we only have limited spectroscopic data available. 
Can we still correct for the bias in a reliable way, using the information
contained in the spectroscopic ``calibration'' set?

There are essentially two nontrivial issues to this end. 
As we pointed out in Rossi \& Sheth (2008), one concern is as to whether or not the number of spectra
which must be taken to specify the error distribution reliably is
sufficient to also provide a reliable spectroscopic estimate of
these fundamental distributions and scaling relations. In this case, the basis for deciding that
it is worth reconstructing these relations from photo-$z$ data is
not clear. However, as long as the spectroscopic sample
spans the \textit{entire} range of photometric observables,  
we show in the next subsection that a detailed knowledge of
the photo-$z$ error distribution (Figure \ref{test_number_spectra_pdf_redshift}) can be inferred
with only $10\%$ of randomly spaced spectra in redshift space. This
rather conservative choice will guarantee an accurate reconstruction of the intrinsic relations. 
Cross-correlations with other surveys may also provide enough reliable information to
specify $p(\zeta|z)$ accurately. 

The second problem is more challenging. If the spectra \textit{are not} simply a random subset of the magnitude
limited photometric sample, then it may be difficult to quantify
and so correct for the selection effects associated with the
spectroscopic subset. In particular, if the spectroscopic sub-sample
\textit{does not} span the entire range of photometric observables, 
it is problematic to perform the reconstruction. 
However, in this situation one may rely 
on photo-$z$ error estimates for the range where spectral
information is missing, and still apply the deconvolution
procedure. We investigate this idea further in the second subsection,
and discuss its applicability and limitations.


\subsection{How many spectra?}

\begin{figure*}
\centering
\includegraphics[angle=0,width=1.03\textwidth]{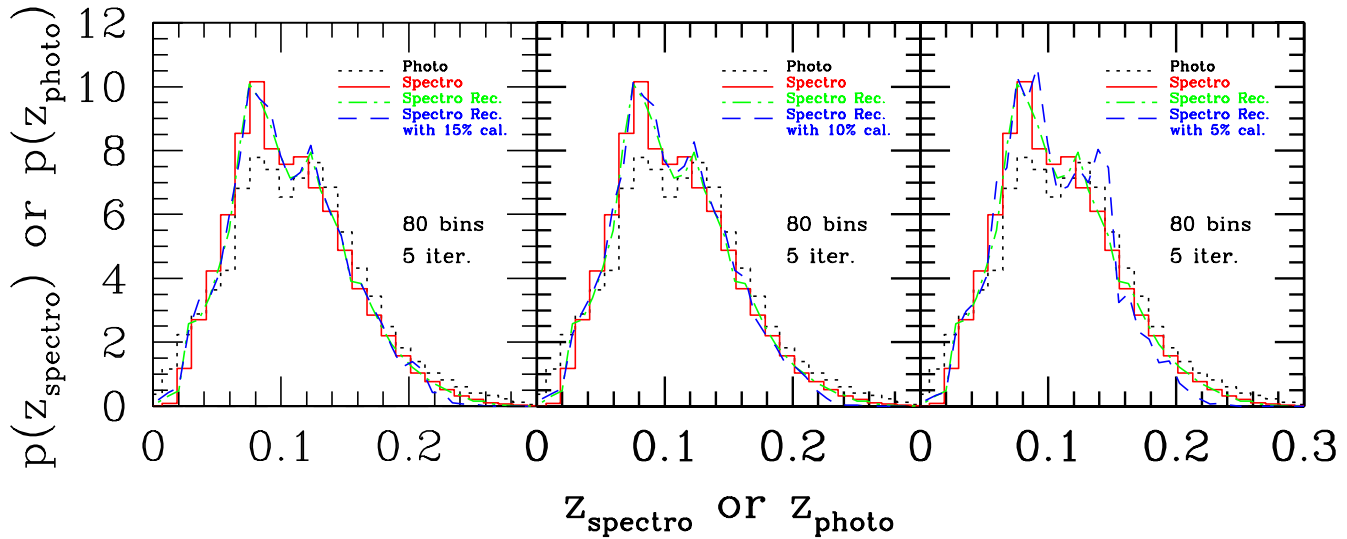}
\caption{Reconstructions of the intrinsic redshift distribution
    (long-dashed lines) when
    only $15\%$ [left panel], 
    $10\%$ [middle panel], or $5\%$ [right panel]  of the original spectral information is used. For comparisons,
    the reconstruction performed with the full early-type spectroscopic information
    is also plotted (``fiducial'' distribution), with jagged
    dot-dashed lines. The solid and dotted histograms are the
    redshift distributions of the early-type spectroscopic and photometric
    catalogs, respectively.}
\label{test_number_spectra_redshift}
\end{figure*}

The accuracy in the reconstruction of the intrinsic redshift 
distribution depends both on the quality of photometric redshifts,
and on the size of the calibration sample with spectral information. 
To study how this accuracy depends on
the size of the calibration sample, we
consider the early-type ``calibration'' catalog and randomly
remove $85\%$, $90\%$, or $95\%$ of the available spectroscopic
information, respectively. 
By this we mean that we are picking random $15\%$, $10\%$ or $5\%$
from the apparent magnitude limited sample.
We refer to this part as the ``degradation'' of the catalog.
We then measure the $p(\zeta|z)$'s conditional distributions for five different
spectroscopic redshift bins, and compare them with those computed when
the full spectroscopic information is available. Figure \ref{test_number_spectra_pdf_redshift}
is the result of this test. Dotted lines in all the panels are
the error probability functions measured in different bins when all the spectral information is
available; long-dashed lines represent the
cases when the catalog is degraded to $15\%$, $10\%$, or $5\%$ of its
original size.   

\begin{figure}
\centering
\includegraphics[angle=0,width=0.50\textwidth]{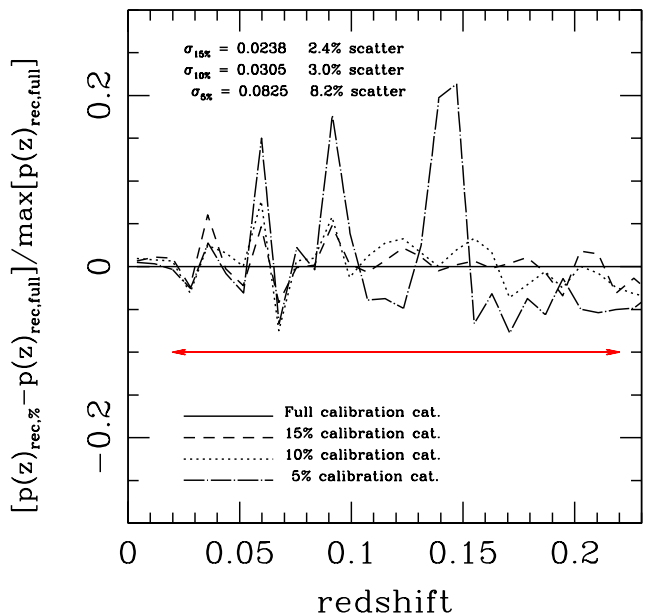}
\caption{Convergence study of the reconstructed solutions
  presented in Figure \ref{test_number_spectra_redshift}, for
  different spectroscopic ``degradation'' levels.
  The root mean square fluctuations of the
  difference expressed in the $y$-axis are also given, in the redshift
  range $0.02 \le z \le 0.22$ denoted by the horizontal arrow, 
  as explained in the main text.}
\label{test_number_spectra_redshift_scatter}
\end{figure}

We now ask how accurately we can reconstruct the intrinsic redshift
distribution using these sub-samples, when the error in the photometric
redshift is given as in Figure \ref{test_number_spectra_pdf_redshift} (the root-mean-square (RMS)
$\delta z = z_{\rm photo}-z_{\rm spectro}$ is typically $\simeq 0.038$).
To this end, we use those ``degraded'' error probabilities to recover the
intrinsic redshift
distribution from photometric data, the result of which is displayed in Figure
\ref{test_number_spectra_redshift}.
Jagged dot-dashed lines show the reconstruction when the full catalog
is used, and long-dashed lines are results of the deconvolutions performed with
limited random spectroscopic subsets.
We quantify the scatter/convergence of the deconvolution procedure in Figure
\ref{test_number_spectra_redshift_scatter}, where we plot the difference
(within each bin) between the reconstructed intrinsic distributions
obtained when  
partial versus full spectroscopic information (i.e. ``fiducial'' distribution) is used, normalized by the
maximum value of the reconstructed fiducial distribution.
Within the figure, we provide the RMS fluctuations of the
difference, in the redshift range $0.02 \le z \le 0.22$ marked by the horizontal arrow in the panel. 
We find that a safe and reliable
reconstruction is guaranteed when the spectral information is
restricted up to $10\%$ of its original size (i.e. $\sim 3\%$ scatter), for randomly
spaced data in redshift space. 

\subsection{Can we use Gaussian approximations?} \label{gauss_prox}

\begin{figure*}
\centering
\includegraphics[angle=0,width=1.05\textwidth]{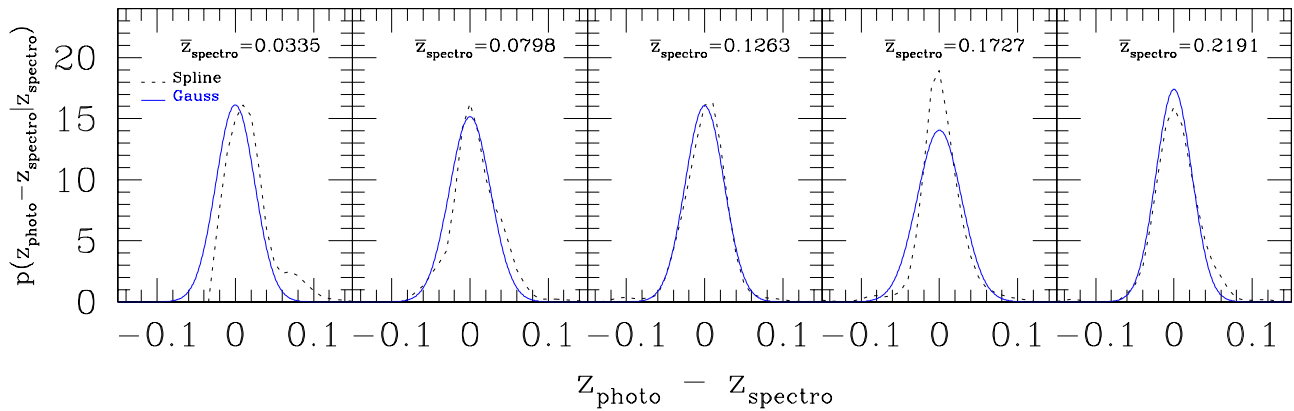}
\caption{Measurements of the conditional probabilities $p(\zeta-z|z)$'s for five different
  redshift bins, as indicated in the panels. Dotted
  lines are spline fits to the pdf's (the full early-type catalog 
  is used), solid lines are unbiased Gaussian approximations
  with widths determined by quadratically averaging
  the SDSS photo-$z$ quoted errors within each redshift bin.} 
\label{test_gauss_proxy_error}
\end{figure*}

\begin{figure}
\centering
\includegraphics[angle=0,width=0.48\textwidth]{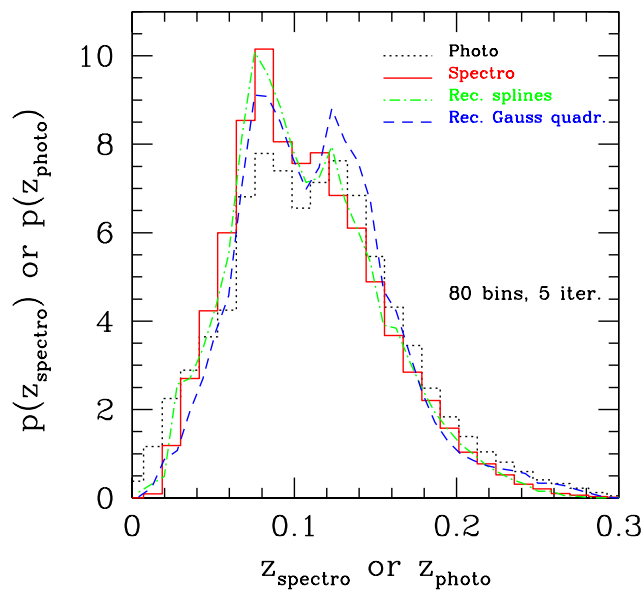}
\caption{Reconstruction of the redshift distribution when Gaussian
  approximations are used for the pdf's (long-dashed line), as opposed
  to the case when the pdf's are measured directly from the data
  (jagged dot-dashed line).}
\label{test_gauss_proxy_redshift}
\end{figure}

Suppose now that we are missing spectroscopic information in some
redshift interval, but that we have
photometry available along with photo-$z$ quoted errors in that range. 
Can we still use our deconvolution technique? In principle, our
method is always readily applicable, provided the knowledge of
$p(\zeta|z)$. How can we infer it, given the lack of spectral
information? 
In effect, the reconstruction of the intrinsic redshift distribution depends not only
on the calibration sample size and on the size of the photometric redshift
error $\delta z$, but also on the shape of the distribution $\delta z$.
Without relying on other surveys which may cover
missing spectral area, we would need to make some
assumptions on its shape. The easiest solution is to derive the conditional distributions $p(\zeta|z)$'s
by using quoted photo-$z$ errors. 
Specifically, one may assume the $p(\zeta|z)$'s to be unbiased
(i.e. $\bar{\zeta} \equiv \bar{z}$) Gaussians, with
widths $\sigma_{\rm z,i}$ determined by quadratically averaging
the SDSS photo-$z$ quoted errors within each redshift bin. 
The question arises as to whether this is a good approximation or not, 
since there is no a priori reason
for the error distributions to be Gaussian and unbiased (Oyaizu et al. 2008). 
We test this idea on the early-type ``calibration'' catalog, and
the result is displayed in Figure \ref{test_gauss_proxy_error}.
In each panel, we show with dotted lines
spline fits to the error probability functions
measured from the early-type catalog (all the spectral information is used in this case), and
with solid lines their unbiased Gaussian approximations.
As it appears evident from the figure, Gaussian approximations are
almost always reasonable fits to the data (excluding small tail
departures or catastrophic photo-$z$ failures), but with the exception
of the redshift interval $z_{\rm
spectro} = 0.1727$, where the Gaussian fit is rather poor. Unfortunately, this departure is sufficient to make the overall
reconstruction of the intrinsic redshift distribution problematic. In
Figure \ref{test_gauss_proxy_redshift} we show, with long-dashed
lines, the outcome of the deconvolution algorithm when unbiased
Gaussian fits are assumed for the pdf's, and report again for
comparison the accurate reconstruction (jagged dot-dashed line), as described in Section \ref{red}.
The deconvolution is critical particularly in the redshift
interval where the Gaussian approximation fails (i.e. around $z_{\rm
spectro} = 0.1727$), and the overall result is of a poor 
recovery of the intrinsic relation.   
Therefore, an accurate knowledge of 
the distribution of photo-$z$ errors is essential for our method to
work.

Nevertheless, improving photo-$z$ uncertainties may help to
characterize $p(\zeta|z)$ more accurately.
It is also worth noticing that, in complete absence of
spectroscopic counterpart, we are in principle still able to apply our technique, 
provided that we have a physically motivated model for the conditional error
distribution. This is the real power of this method. 
On the opposite,  the weighting technique proposed by Lima et al. (2008),
which addresses similar goals, always requires the spectroscopic 
sub-sample to span the \textit{entire} range of
photometric observables covered by the photometric sample.



\section{Discussion}

Using a selected sample of early-type galaxies from the SDSS DR6, for which both photo-$z$s and
spectro-$z$s are known, we applied our one- and two-dimensional deconvolution 
techniques (Sheth 2007; Rossi \& Sheth 2008) to reconstruct the
unbiased redshift, magnitude and size distributions, as well as
the magnitude-size relation (Section \ref{reconstr}).  

This is a novel approach in recognizing that
theoretical predictions, such as the difference between photometric
and spectroscopic distance estimates, can be represented as integral
equations and solved using deconvolution techniques. In the past,
these techniques have been used only observationally, for instance in
handling the PSF of telescopes. 

We showed that our technique recovers all the true distributions and
the joint relation, to a good degree of accuracy.
We discussed the magnitude dependence of the error conditional
probabilities (Section \ref{red}), and argued that the problem of reconstructing the true magnitude or size distribution
is best thought of as a one-dimensional deconvolution problem
(Sections \ref{mag_dist} and \ref{size_dist}).  
We showed that even if the distribution of physical sizes is not
severely biased, a significant bias in the magnitude distribution suffices to compromise the
size-luminosity relation (Section \ref{size_lum_corr}). 
We used our 2D deconvolution technique to correct for this effect.

We then discuss how to extend our procedure
to deep redshift catalogs, where limited spectroscopic information, or
only photometric data, is
available (Section \ref{ext}). For this part, we performed two tests
using the early-type ``calibration'' sample. 
We found that using only $10\%$ of the spectroscopic information randomly spaced
in our catalog is sufficient for the reconstructions to be accurate
with about $3\%$ scatter, when the error in the photometric redshift is
typically $\delta z \simeq 0.038$.
We also showed that assuming unbiased Gaussians for the
$p(\zeta|z)$'s distributions, with widths determined by quadratically averaging
the SDSS photo-$z$ quoted errors within each redshift bin, is not always a good
approximation. However we argued
that, provided one has a detailed knowledge of the pdf from other
surveys or from empirically motivated models (see for instance van der
Wel et al. 2009), our technique can still
be used, even when the spectroscopic sample
does not span the entire range of
photometric observables covered by the photometric sample.

We address in more detail the problem of
handling photo-$z$s when spectroscopic information is missing (for
example using a ``blind'' deconvolution approach) in a
forthcoming publication, where we also apply our technique to 
reconstruct the luminosity function in deep redshift catalogs such as
the MegaZ-LRG (Collister et al. 2007).  

Even though our discussion was mainly phrased in terms of fundamental
distributions and scaling relations for 
early-type galaxies, so it may be useful for detailed studies of
early-types (for example van den Bosch \& van de
Ven 2008; Bernardi 2009), the method developed here is quite general and
can be applied to recover any intrinsic correlations between
distance-dependent quantities (even for $n$-correlated variables);
potentially, it can impact a broader range of studies, when at least
one distance-dependent quantity is involved.
In fact, our algorithms can be readily adapted to study the
luminosity function in relatively shallow peculiar velocity surveys
with noisy Fundamental Plane or $D_{\rm n}-\sigma$ distance estimates
(Faber et al. 2007; Tully et al. 2009),
or to handle correctly uncertainties in supernova measurements (Krauss et
al. 2007; Frieman et al. 2008; Sako et al. 2008), which can bias the redshift-dependent equation of state (Bridle
\& King 2007; Fosalba \& Dore 2007). 

A variety of other correlations
can be re-analyzed along these lines (see for example Saracco et
al. 2009), 
such as the $R-L$ relation for blue
galaxies (Melbourne et al. 2007), the photometric Fundamental Plane
(Bolton et al. 2007), and also correlations that do not involve
luminosity, such as the Kormendy (1977) relation.Other possible applications involve quasars (Croom et al. 2009; Richards et al. 2009), black-hole $M-L$
correlations, correlations with environment,
and potentially future baryonic acoustic oscillation and dark energy surveys.



\section*{Acknowledgments}

We thank an anonymous referee for useful comments and suggestions.
GR would like to thank Bruce Bassett and Penjie Zhang for stimulating discussions about
deconvolution techniques. 
CBP acknowledges the support of the Korea Science and Engineering
Foundation (KOSEF) through the Astrophysical Research Center for the
Structure and Evolution of the Cosmos (ARCSEC).

Funding for the SDSS and SDSS-II has been provided by the Alfred P. 
Sloan Foundation, the Participating Institutions, the National Science 
Foundation, the U.S. Department of Energy, the National Aeronautics and
 Space Administration, the Japanese Monbukagakusho, the Max Planck 
Society, and the Higher Education Funding Council for England. The 
SDSS Web Site is {\tt http://www.sdss.org/}.

The SDSS is managed by the Astrophysical Research Consortium for the
Participating Institutions. The Participating Institutions are the
American Museum of Natural History, Astrophysical Institute Potsdam,
University of Basel, University of Cambridge, Case Western Reserve
University, University of Chicago, Drexel University, Fermilab, the
Institute for Advanced Study, the Japan Participation Group, Johns
Hopkins University, the Joint Institute for Nuclear Astrophysics,
the Kavli Institute for Particle Astrophysics and Cosmology, the 
Korean Scientist Group, the Chinese Academy of Sciences (LAMOST), 
Los Alamos National Laboratory, the Max-Planck-Institute for Astronomy 
(MPIA), the Max-Planck-Institute for Astrophysics (MPA), New Mexico 
State University, Ohio State University, University of Pittsburgh, 
University of Portsmouth, Princeton University, the United States 
Naval Observatory, and the University of Washington. 




\appendix

\section{Direct Reconstructions}

A widely used method for extracting the redshift
probability distribution function (PDFz) from a $\chi^2$ minimization
is the following (see for example Bolzonella et al. 2000). 
A $\chi^2$ is formed, typically as
\begin{equation} 
\chi^2 = \sum_{\rm f=1}^{\rm N_f} 
\Big ( 
{F_{\rm obs}^{\rm f} - A \cdot
  F_{\rm pred}^{\rm f}(z,T) \cdot 10^{-0.4 s_{\rm f}} \over
 \sigma_{\rm obs}^{\rm f}}
 \Big )^2
\label{chi_sq} 
\end{equation} 
\noindent where $F_{\rm pred}^{f}(z,T)$ is the flux
predicted for a template T at redshift z, $F_{\rm
obs}^{\rm f}$ is the observed flux, $\sigma_{\rm
obs}^{\rm f}$ is the associated error, $f$ refers to each
specific filter and $s_{\rm f}$ is the zero-point offset.
The photo-$z$ is estimated from the $\chi^2$ minimization
with respect to the free parameters z, T, and
the normalization factor A; namely, the photo-$z$ is the redshift
value which minimizes the merit function $\chi^2(z,T,A)$.
The associated PDFz, or $p(z|\zeta$), is derived
from (\ref{chi_sq}),
\begin{equation} 
  p(z|\zeta) \propto e^{-[\chi^2(z)-\chi^2_{\rm
        min}(\zeta)]/2}\,. 
  \label{pz_given_zeta} 
\end{equation} 

There is a main conceptual point in adopting this approach. Photo-$z$s are noisy
distance estimates, as opposed to  spectroscopic
redshifts, which are intrinsic or ``true''
solutions. Therefore, while $\langle
z_{\rm photo}|z_{\rm spectro} \rangle \rightarrow z_{\rm
spectro}$ (i.e. convergence of the noisy distribution to
the true value), it is certainly not true that $\langle
z_{\rm spectro}|z_{\rm photo} \rangle \rightarrow z_{\rm
photo}$. This is equivalent to say that the distribution $p(z|\zeta)$, obtained by 
binning horizontally the plane [$z_{\rm photo}$, $z_{\rm spectro}$] shown in the
left panel of Figure \ref{redshift_reconstruction_early_types}, 
is biased by definition.
Hence, it is  more meaningful to estimate
$p(\zeta|z)$ rather than attempting to derive
$p(z|\zeta)$. 

However, since current photo-$z$ codes output 
$p(z|\zeta)$ rather than $p(\zeta|z)$, one may wonder if 
we can apply our technique using the PDFz.
In effect, our deconvolution method relies on
Bayes's theorem. For example, if we consider the redshift
distribution, application of this theorem yields
\begin{equation}
  p(z,\zeta) = N(z) \cdot p(\zeta|z) \equiv p(\zeta, z)
  = {\cal N}(\zeta) \cdot p(z|\zeta).   
\end{equation}  
From the previous relation, it is immediate to show that
\begin{equation}
  N(z) = \int {\cal N}(\zeta)~p(z|\zeta)~{\rm  d}\zeta\,.   
  \label{no_deconvolution}
\end{equation} 
In this respect, the true (spectroscopic) redshift
distribution can be alternatively viewed as a convolution of the noisy photo-$z$
distribution times the PDFz. Therefore, assuming that 
the PDFz is known from the output of  photometric redshift codes and given the observed photo-$z$
distribution, one can obtain the intrinsic $N(z)$ by
simply performing the integration (\ref{no_deconvolution}).  
This idea is further explored in Sheth \& Rossi (2009), where 
examples of these calculations are presented using the SDSS sample
described here. 

Similarly, one can obtain the magnitude distribution
with a direct integration, since
\begin{equation}
  N(M) = \int {\cal N}({\cal M})~p(M|{\cal M})~{\rm  d}{\cal M}\,,   
  \label{no_deconvolution_mag}
\end{equation} 
and in principle recover scaling relations
as well. However, we would like to remind the reader that when dealing
with a real dataset, a noisy observation needs to be ``deconvolved'' into an intrinsic
signal, namely from ${\cal N}(\zeta)$ one needs to
reconstruct $N(z)$; $p(z|\zeta)$ is usually not known while
$p(\zeta|z)$ can be inferred reliably with a proper
``spectroscopic training set'' -- this is our main motivation for providing a deconvolution
approach. 


\label{lastpage}
\end{document}